\documentclass[aps,prb,twocolumn,showpacs,preprintnumbers,amsmath,amssymb]{revtex4}
\usepackage{graphicx}
\usepackage{hyperref}
\usepackage{natbib}
\usepackage{dcolumn}
\usepackage{bm}

\begin{document}

\title{Tuning the gap in bilyaer graphene using chemical
           functionalization: DFT calculations}

\author{D. W. Boukhvalov}
\email{D.Bukhvalov@science.ru.nl} \affiliation{Institute for
Molecules and Materials, Radboud University of Nijmegen, NL-6525
ED Nijmegen, the Netherlands}
\author{M. I. Katsnelson}
\affiliation{Institute for Molecules and Materials, Radboud
University of Nijmegen, NL-6525 ED Nijmegen, the Netherlands}

\date{\today}

\begin{abstract}
Opening, in a controllable way, the energy gap in the electronic
spectrum of graphene is necessary for many potential applications, 
including an efficient carbon-based transistor. We have shown that this can
be achieved by chemical functionalization of bilayer graphene.
Using various dopants, such as H, F, Cl, Br, OH, CN, CCH, NH$_2$,
COOH, and CH$_3$ one can vary the gap smoothly between 0.64
and 3 eV and the state with the energy gap is stable corresponding
to the lowest-energy configurations. The peculiarities of the structural
properties of bilayer graphene in comparison with bulk graphite
are discussed.
\end{abstract}

\pacs{73.20.Hb, 71.15.Nc, 81.05.Uw, 61.46.Np, 61.48.De, 72.80.-r}

\maketitle

\section{Introduction}

Graphene, a recently discovered two-dimensional allotrope of
carbon (for review, see Refs.\onlinecite{r1,r2,r3}) is a very
promising material for future development of electronics, due to
its planar geometry and a very high electron mobility \cite{r1}.
Investigations of graphene create a new, unexpected bridge between
condensed matter physics and quantum electrodynamics (for review,
see Ref.\onlinecite{r4}). At the same time, some of the exotic quantum
phenomena which make graphene so attractive scientifically can be
considered as obstacles for applications. In particular, the
chiral ``Klein'' tunneling \cite{klein} makes $p-n-p$ (or $n-p-n$)
junctions unusually transparent. This does not allow to lock the
junction making its use as a transistor problematic. Bilayer
graphene \cite{natphys} is 
preferable in this sense since the
angular range of anomalous transparency is narrower there
\cite{klein} but only the opening of a real gap in electron spectrum
would be a radical solution of the problem. The gapless conical
spectrum in the single-layer graphene is very robust; actually, it
is protected topologically, assuming that one does not break the
sublattice equivalence \cite{guinea}. The latter can be done,
e.g., in a hypothetic bilayer system consisting of a single-layer
graphene and a single-layer hexagonal boron nitride \cite{BN} but
the gap which can be opened in this way is rather small, only about 50
meV. The robustness of the gapless state in the single-layer graphene was
demonstrated in recent electronic structure calculations for
hydrogenated graphene \cite{H}. It turns out that the gap opens in
this case only at 75\% coverage. In bilayer graphene, the gap
can be opened by applying a strong electric field perpendicular to
the graphene plane, as it was demonstrated by recent experimental
\cite{gap1,gap2} and theoretical \cite{gap3,gap4} investigations.
In this case the gap is tunable but, again, only in some
restricted limits, not larger than the middle infrared region.

Chemical modification of bilayer graphene seems to be a
natural way to tune the gap in broader range, from zero to
the values typical for conventional semiconductors such as silicon
or GaAs. This is the subject of the present work. It is shown
that, in contrast with the case of single-layer graphene, the gap opens for
dopant concentrations corresponding to the most stable
configuration.

\section{Computational method}

We used the SIESTA package for electronic structure calculations
\cite{siesta1,siesta2} with the generalized gradient approximation
for the density functional \cite{perdew}, with the energy mesh cutoff
of 400 Ry, and a $k$-point 11$\times$11$\times$1 mesh in the Monkhorst-Park
scheme \cite{MP}. This method is frequently used for computations
of the electronic structure of single-layer graphene
\cite{H,Yazyev,Louie,Kaxiras}.

It is known \cite{Janotti,Simak,Marini} that the use of GGA leads
to essential overestimate of the equilibrium interlayer distances for
layered compounds, such as graphite, hexagonal boron nitride
(hBN), and MoS$_2$, due to the inadequate description of the van der
Waals interaction effects. On the other hand, the LDA slightly
underestimates these distances \cite{Janotti,lapw}. However, all
these calculations have been carried out for three-dimensional
crystals where each layer interacts with {\it two} neighboring
layers and it is not clear {\it a priori} what is the situation
for a bilayer. Structural properties of bilayer graphene are
different from both bulk graphite and single-layer graphene, as is
confirmed by measurements of the Raman spectra \cite{Ferrari} and
characteristics of ripples on the bilayer \cite{Meyer}.
Peculiarities of structural states of bilayers in comparison with
bulk crystals were observed also in ionic crystals such as
wurtzite ZnO(0001) \cite{ZnO}.

\begin{figure}
\rotatebox{-90}{
\includegraphics[height=3.2 in]{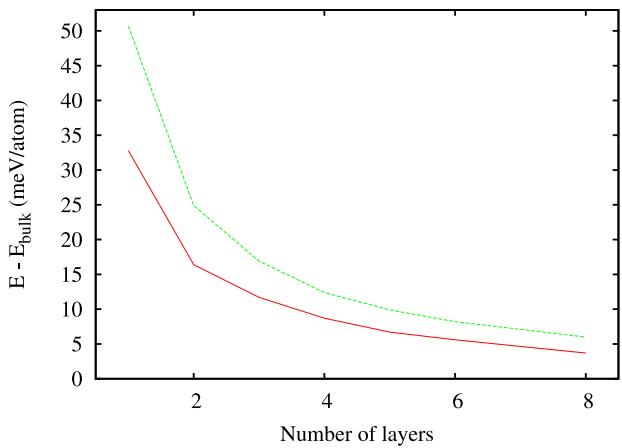}}
\caption{\label{fig1} (color online) Total energies per atom for
multilayer graphene (solid red line) and hexagonal BN (dashed
green line) counted from those for bulk, as functions of number of
layers.}
\end{figure}

To check the applicability of different approximations we have carried
out LDA and GGA calculations of structural properties of the
multilayers of graphene and hBN. The LDA computational results
are shown in Fig. \ref{fig1}. One can see that the total energy per
atom almost coincides with that for bulk graphite starting with
approximately five layers (with the accuracy of 1 meV), in 
qualitative agreement with the conclusions from Raman spectra
\cite{Ferrari}. The energy difference between bilayer and single-layer 
is approximately the same as between the bulk crystal and
bilayer, both for carbon and for BN.

\begin{figure}
\includegraphics[height=3.2 in]{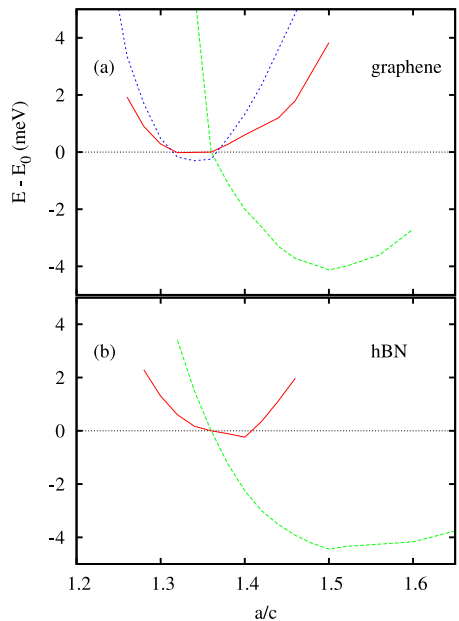}
\caption{\label{fig2} (color online) Total energies per atom as
functions of the ratio of interlayer distance $c$ to in-plane
lattice constant $a$ for graphene (a) and hexagonal BN (b); solid
red and dashed green lines correspond to LDA and GGA,
respectively, dotted blue line corresponds to the LDA calculations
for bulk graphite. Here $E_0$ is the energy of bilayer with
lattice parameters corresponding to the bulk.}
\end{figure}

Interestingly, the equilibrium interlayer distances for bilayer
graphene and hBN differ only within 10\% (Fig. \ref{fig2}) which
is much smaller than for the bulk \cite{Janotti}. Our SIESTA
computational results for the latter case coincides within 1\%
with those of FLAPW calculations \cite{Janotti} so this difference
is rather due to the difference of the systems themselves than of the
methods used. It is worthwhile to note that the curves of energy
versus interlayer distance are essentially different for the bulk
and for the bilayer.

Thus, LDA and GGA computational results are much closer for the
case of bilayers than for the case of bulk layered crystals.
In the following, we will discuss in more detail only the GGA data since GGA
gives better results for chemisorption energies and bond lengths
\cite{H}. Anyway, neither LDA nor GGA allows to calculate
accurately energy gaps \cite{yang} and the corresponding results
are rather estimations from below. At the same time, as we will
see, that the energy gap grows as the interlayer distance decrease so
the GGA overestimating interlayer distances gives surely the lower
estimation whereas, in the case of LDA, the error can be,
in principle, of any sign.

\begin{figure}
\includegraphics[height=3.2 in]{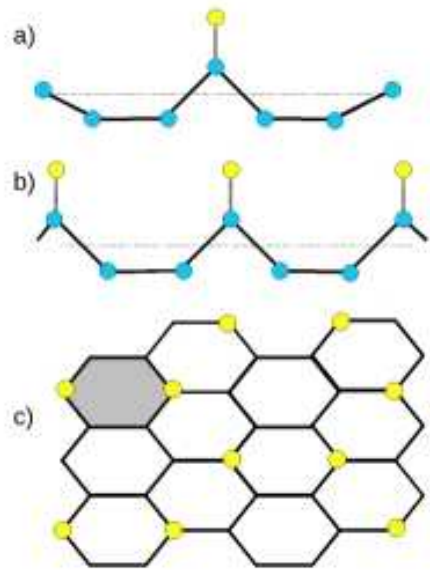}
\caption{\label{fig3} (color online)
A sketch of atomic positions for hydrogenated graphene:
(a) A single hydrogen atom; (b) the most stable configuration for
the chemisorption at one side; (c) the maximum coverage for the
chemisorption at one side (a top view). Blue circles are carbon
atoms, yellow ones are hydrogens, the grey hexagon corresponds to
the atomic group shown in Figs. \ref{fig4} and \ref{fig6}.}
\end{figure}

Recent electronic structure calculations for hydrogenized graphene
\cite{H} demonstrate that it is more favorable energetically to
attach dopants to carbon atoms belonging to different sublattices,
which allows to avoid the formation of dangling bonds. Minimization of
geometric frustration of the carbon lattice is another important
factor determining the stable configurations. Fig. \ref{fig3}a
displays schematically the distortion of single-layer graphene
for chemisorption of single hydrogen atom: the carbon atom
connected with hydrogen is shifted up whereas its nearest and
next-nearest neighbors are displaced down and the third neighbors
are shifted up again. The most stable configuration for the case
of single-layer graphene correspond to bonding of hydrogen with
neighboring carbon atoms at opposite sides (positions 1 and 2,
according to the standard chemical terminology). If one allows
only the one-side chemisorption than (1,4) positions of hydrogens
turn out to be optimal, that is, bonding with third neighbors
(Fig. \ref{fig3}b). In that case distortions of positions of other
carbon atoms will be similar to those for (1,2) bonding (nearest
and next nearest neighbors are shifted down). For the
bilayer only one side of each graphene layer is available so one
can expect an optimal configuration similar to shown in Fig.
\ref{fig3}b. One can expect that these arguments are applicable
not only to hydrogen but to other dopants, and this is confirmed,
indeed, by our computational results.

This leads to the important consequence that the maximum coverage for
the bilayer should be about 25\%, otherwise first and
second neighbors are filled unavoidably. Contrary, for the case of
single-layer graphene where both sides are available the
configuration with one hydrogen atom per carbon atom is the most
energetically favorable \cite{H}. The optimal one-side coverage of
bilayer is sketched in Fig. \ref{fig3}c.

\section{Results and discussions}

\begin{figure}
\includegraphics[width=3.2 in]{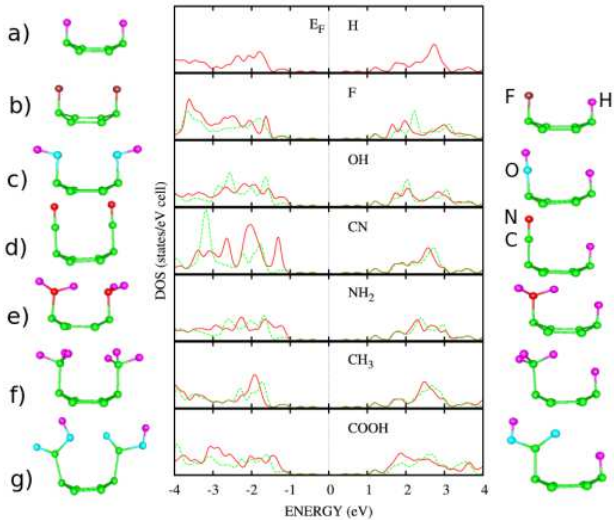}
\caption{\label{fig4} (color online)
Optimized configurations and total densities of states
for one-side functionalization of bilayer graphene. Left panel and
red solid lines correspond to the case of two identical dopants,
e.g., F...F, per hexagon; right panel and dashed green lines
correspond to the case when one dopant group per hexagon is
replaced by hydrogen atom, e.g. F...H.}
\end{figure}

\begin{figure}
\includegraphics[height=4.2 in]{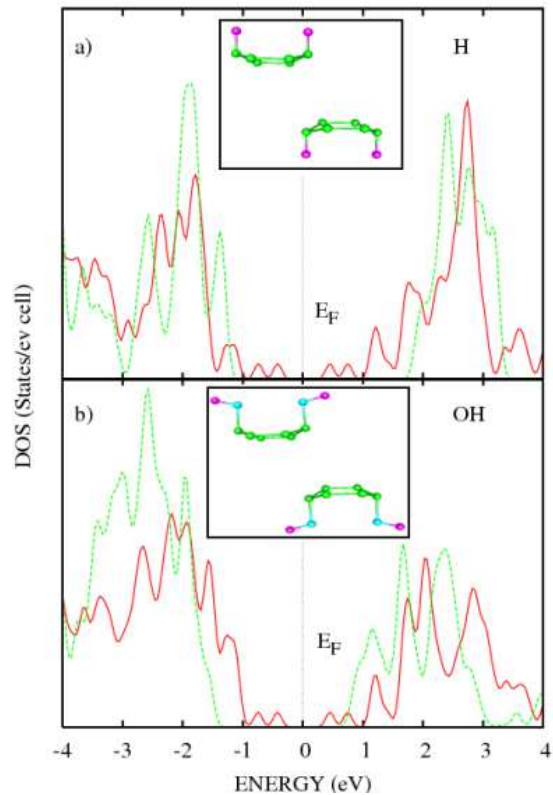}
\caption{\label{fig5} (color online)
Total densities of states for one-side (solid red lines)
and two-side (dashed green lines) functionalization of bilayer
graphene, for the case of hydrogen (a) and hydroxyl (b). Insets
show optimized atomic configurations for the case of two-side
functionalization.}
\end{figure}

First, we have investigated the dependence of the total energy on
the type of dopant and its concentration (coverage level). We have
found that the most stable configurations occurs for functionalization by
fluorine and hydroxyl groups (the case of hydrogen was considered
in Ref.\cite{H}). The supercell size was varied between 32 to 8
carbon atoms per layer (the latter corresponds to a maximum possible
25\% coverage, two dopant atoms or molecules per eight carbon
atoms). The chemisorption energy was calculated as described in
Ref.\onlinecite{H}, choosing molecular fluorine F$_{2}$ and water
H$_{2}$O as reference points for the cases of F and OH,
respectively. The calculations show a decrease of the
chemisorption energy with coverage (in the limits
noticed above), from - 1.58 eV to - 1.89 eV for F and from -2.37
eV to -3.16 eV for OH. In both cases, as well as for H, the most
stable configurations correspond to the maximum coverage. While 
for single H atom the activation energy is positive (1.28 eV) and
thus its chemisorption is not favorable, for F and OH the
corresponding values are -1.21 eV and -2.23 eV, respectively.

Further, we have considered also the one-side functionalization of
bilayer graphene by other groups, namely, CN, NH$_2$, CH$_3$,
COOH, as well as by combination of the dopants and hydrogen (see
Fig. \ref{fig4}). Despite a rather different chemical composition
of dopants the distortions of the functionalized graphene layer
the height differences between the highest and the lowest positions
of carbon atoms in the layer $d$ lie in a relatively narrow
interval, from 0.36 \AA ~for the case of H to 0.57 \AA ~for the case
of COOH (see Fig. \ref{fig4}g), that is, from 11\% to 17\% of
interlayer distance in graphite (3.35 \AA). The minimal interlayer distances
$h$ in the one-side functionalized bilayer vary between 3.25 \AA
~for H and 2.98 \AA ~for COOH (97\%-89\% of interlayer distance in
graphite).

\begin{figure}
\rotatebox{-90}{
\includegraphics[height=3.2 in]{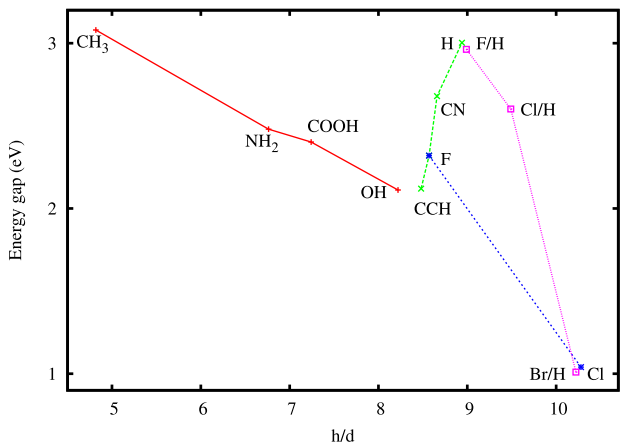}}
\caption{\label{fig6} (color online)
Energy gap values for the case of two-side
functionalization of bilayer graphene, as a function of the ratio of
the interlayer distance $h$ to the carbon atom distortion $d$.}
\end{figure}

The electronic structure for the most stable configuration (25\%
coverage) is shown in Figs. \ref{fig4} and \ref{fig5}. One can see
that, surprisingly, the value of the energy gap opening as a function of
doping is not too sensitive to the dopant type varying in the
interval 0.64 - 0.68 eV (see Fig. \ref{fig6}). In a sense, the
situation reminds the epitaxial graphene on SiC \cite{epitax}
where one graphene layer is supposed to be almost unperturbed and
another one (buffer layer) is strongly coupled covalently with the
substrate. For some types of epitaxial graphene, the existence
of energy gap was theoretically predicted (0.45 eV,
Ref.\onlinecite{gapth}) and experimentally confirmed (0.26 eV,
Ref.\onlinecite{gapexp}). Interestingly, some peaks of the density of
states around the Fermi energy have been observed there
\cite{gapexp}. Similar peaks can be seen also in our computational
results (Fig. \ref{fig4} and \ref{fig5}).

Further, we have investigated the case of two-side
functionalization. It was shown already that for the case of
hydrogen, one-side and two-side chemisorption energies of bilayer
graphene are rather close \cite{H}. The same turns out to be
the case also for the case of fluorine and hydroxyl group. At the
same time, the electronic structures for the case of one-side and
two-side doping are completely different. For the latter case, the
energy gap is essentially larger varying from 2.12 eV for hydroxyl
group to 3.03 eV for hydrogen (see Fig. \ref{fig5}). In the case
of one-side functionalization a hybridization with practically
unperturbed layer of pure graphene holds whereas for the two-side
case both layers are strongly modified and distorted.

We now discuss the equilibrium configurations of the dopants shown
in Figs. \ref{fig4} and \ref{fig5}. For single atoms such as H and
F it is impossible to discuss their orientation, and for the case
of CN group a very strong triple C-N bond keeps the molecule
normal to the graphene plane. For other cases, the dopant orientations
are coordinated by their interactions. This is clearly seen, e.g.,
for OH group in Fig. \ref{fig4} and inset to Fig. \ref{fig5}b.
While for a single OH group the angle C-O-H is almost
180$^\circ$ for neighboring OH groups, this angle diminishes to
105$^\circ$ and a preferable mutual orientation of the groups
appear, for both one-side and two-side doping (see inset to Fig.
3b for the latter case). For the dopants of this type, actually, a
four-layer system is formed, such as dopant/carbon/carbon/dopant.
This leads to specific dopant-dependent distortions of graphene
affecting the value of the energy gap. The stronger distortion
and, therefore, the weaker the interaction between the dopants at
opposite sides of graphene, the larger is the energy gap. For
the different single-atom dopants under consideration, the distortions
are more or less the same and the gap is mainly dependent on their
number in the Periodic Table. These data are summarized in Fig.
\ref{fig4}.

We have performed also calculations for CCH dopant, with the
triple C-C bond. In this case, as well as for CN, the dopant
orientation is irrelevant, and the value of the gap continues the
line H-CN-F-... (Fig. \ref{fig6}).

We have considered also different combinations of the dopants with
hydrogen. The latter destroys the ordered four-layer structure
described above and the values of energy gap turn out to be close
for all the dopants varying in the limits 2.96 - 3.03 eV.

Next, we have investigated the effect of surrounding water for the
case of hydroxyl groups. If one adds one water molecule per group
connected by the hydrogen bond and optimize the structure it leads
to additional distortions of the bilayer (0.39 \AA ~without water
and 0.51 \AA ~with water) increasing the value of energy gap from
2.12 to 2.36 eV.

We have considered also the doping of bilayer by heavier elements,
namely, halogens Cl and Br. However, in these cases the
limitations because of size factors become more essential. Whereas
doping of the bilayer by chlorine is possible two bromine atoms
form stable molecule B$_2$ under the surface of practically
undistorted bilayer. However, partial doping by heavy halogens is
possible if combine them with hydrogen. The values of energy gap
for F...H and Cl...H doping are larger than for F...F and Cl...Cl,
respectively (see Fig. \ref{fig6}).

\section{Conclusion}

To conclude, our results allow to formulate general principles
determining the value of energy gaps in doped bilayer graphene.
One-side doping, almost independently on the chemical nature of
the dopants, leads to gaps of the order of 0.6 - 0.7 eV. Two-side
doping makes possible to change the gap in much broader limits.
The functionalization by halogens or their combination with
hydrogen results in gap values in the range 1 to 3 eV,
however, this value cannot be fine tuned. On the other hand, using
various groups formed by elements of the second period of the
Periodic table one can change the gap between 2 and 3 eV
smoothly, with the accuracy about 0.2 eV. Interaction between
dopants and water can also change the gap by a value of order of
0.2 eV. Thus, variations of solvents can be also used to tune the
gap. The case of hydroxyl groups requires further more detailed
investigation due to its relevance for perspective exfoliated
graphene oxide \cite{exfol}.

\section{Acknowledgements}

The work is financially supported by Stichting voor Fundamenteel
Onderzoek der Materie (FOM), the Netherlands.

\end{document}